\documentclass[reprint,amsmath,amssymb,aps,prd,longbibliography,nofootinbib]{revtex4-2}

\usepackage{framed}
\usepackage{amsmath,amssymb,bm}
\usepackage{graphicx}
\usepackage{svg}
\usepackage{tabularx}
\usepackage{ulem}
\usepackage[usenames,dvipsnames,svgnames,table]{xcolor}
\usepackage{enumitem}
\usepackage{comment}
\usepackage{float}
\usepackage{subcaption}
\usepackage{xfp}
\usepackage{dcolumn}
\usepackage{lineno}
\usepackage{booktabs}
\usepackage{bm}
\usepackage[english]{babel}
\usepackage{microtype}
\PassOptionsToPackage{hyphens}{url}
\usepackage{hyperref}

\setlist[itemize]{leftmargin=*}
\graphicspath{{gfx/}} 

\newlength\normalFontSize\newlength\fontSizeIncrement
\def\beq{\begin{equation}}
\def\eeq{\end{equation}}
\def\bea{\begin{eqnarray}}
\def\eea{\end{eqnarray}}

\definecolor{myyellow}{rgb}{0.94, 0.86, 0.51}
\definecolor{mygreen}{rgb}{0.2, 0.8, 0.2}
\definecolor{mypink}{rgb}{0.99, 0, 0.99}
\definecolor{mypurple}{rgb}{0.75, 0, 0.75}
\definecolor{cadmiumorange}{rgb}{0.93, 0.53, 0.18}

\newcommand{\IGNORE}[1]{}

\makeatletter
\providecommand{\href@noop}[0]{\@secondoftwo}
\makeatother

\begin{document}

\preprint{APS/123-QED}

\title{TROYE: Modeling Dynamic Phase Transitions\\in Gravitational Waves from Neutron Star-Black Hole Mergers}

\author{Ofek Dan}
 \email{ofekdan97@gmail.com}
\author{Ofek Birnholtz}
 \email{ofek.birnholtz@biu.ac.il}
\affiliation{Department of Physics, Bar-Ilan University, Ramat-Gan, Israel}

\date{\today}

\begin{abstract}
The Equation of State (EoS) of dense nuclear matter remains one of the most compelling open questions in high-energy astrophysics. While static EoS models are increasingly well-constrained by observations of binary neutron star (BNS) inspirals, the possibility of a dynamic phase transition occurring \textit{during} the coalescence has been thus far deferred from standard gravitational-wave (GW) analyses. In this work, we investigate the detectability of such a phase transition—manifesting as a macroscopic shift in the tidal deformability parameter $\Lambda$—using GWs from Neutron Star-Black Hole (NSBH) coalescences. We argue that NSBH systems serve as a cleaner laboratory for this phenomenology than BNS systems due to the absence of the $\tilde{\Lambda}(\Lambda_1,\Lambda_2)$ degeneracy, allowing for the isolation of single-body tidal evolution. We introduce a phenomenological waveform model, \textsc{troye} (Transitional Representation Of varYing Equation-of-state), which stitches together two waveform approximants in the time domain to simulate a smooth but rapid transition between two equations of state during the late inspiral. We perform a comprehensive Bayesian injection and recovery campaign on 100 simulated events using the \texttt{bilby} inference library. Our results demonstrate that a phase transition corresponding to a tidal shift of $|\Delta\Lambda| \gtrsim 400$ is detectable with Advanced LIGO design sensitivity, yielding decisive statistical evidence ($\ln\mathcal{B} > 5$). We further identify a "V-shape" asymmetry in detectability, where "softening" transitions (decreasing $\Lambda$) are systematically easier to detect than "stiffening" ones due to the specific phase evolution of the tidal sector. Finally, we present "stress tests" showing that the transition remains recoverable even when marginalized over uncertainties in the stitching time and binary mass ratio.
\end{abstract}

\maketitle

\section{\label{sec:intro}Introduction}

Neutron stars (NS) contain the densest matter in the observable universe, offering a unique laboratory for Quantum Chromodynamics (QCD) in regimes inaccessible to terrestrial collider experiments \cite{Lattimer2004, ozel2016}. The macroscopic properties of these stars—their masses, radii, and tidal deformabilities—are governed by the nuclear Equation of State (EoS). While the EoS at saturation density ($\rho_0 \approx 2.8 \times 10^{14}$ g/cm$^3$) is relatively well-constrained by nuclear theory and experiments, the behavior of matter at the supranuclear densities found in NS cores ($ \rho \gtrsim 2\rho_{0}$) remains highly uncertain \cite{baym2018, abbott2018, de2018tidal}.

Theoretical models suggest that at these extreme densities, hadronic matter may undergo phase transitions to exotic degrees of freedom. Possibilities include the formation of hyperonic matter, pion or kaon condensates, or a deconfined quark-gluon plasma (quark matter) \cite{Drago2014, huth2022}. Such transitions are often associated with a "softening" of the EoS, which can lead to the existence of distinct families of compact stars or unstable branches in the mass-radius relation.

Gravitational waves (GWs) from compact binary coalescences provide a direct probe of the NS interior through the measurement of tidal effects. As the binary components spiral closer, the tidal field of the companion induces a quadrupole moment in the NS. This deformation extracts energy from the orbit, accelerating the inspiral and imprinting a phase correction on the GW signal \cite{Flanagan2008, hinderer2008}. To date, standard analyses by the LIGO-Virgo-KAGRA (LVK) collaboration assume a \textit{static} EoS, implying that the tidal deformability $\Lambda$ is a fixed function of the mass throughout the inspiral \cite{abbott170817, abbott2020gw190814}.

However, the environment of a binary inspiral is dynamic. As the orbital separation decreases, the neutron star is subjected to a rapidly strengthening and time-varying tidal field, and may also experience dissipation and heating through mode excitation and related non-adiabatic processes. If the star's internal conditions cross a critical threshold during this evolution, a phase transition to exotic matter could be triggered on dynamical timescales, milliseconds before merger \cite{Drago2014, katerina2020phasetrans}. Such a transition could change the effective stiffness of the star during the inspiral, resulting in a temporal variation of $\Lambda(t)$. We emphasize that this trigger need not be tied to monotonic tidal compression of the core; tidal interactions can modify stellar structure and stability more generally, and can even reduce the central density near marginal stability \cite{thorne1998, lai1996}.

This paper explores the phenomenological detectability of such an event. We focus specifically on Neutron Star-Black Hole (NSBH) systems. While this framework is in principle applicable to Binary Neutron Star (BNS) mergers via the effective tidal parameter $\tilde{\Lambda}$, the interpretation of such a signal is complicated by degeneracy. Since $\tilde{\Lambda}$ is a mass-weighted combination of $\Lambda_1$ and $\Lambda_2$, attributing a temporal shift to a specific component (or distinguishing single from double transitions) is ambiguous. In an NSBH system, assuming the "no-hair" theorem holds for the black hole ($\Lambda_{BH}=0$), all tidal effects originate from the single NS. This breaks the degeneracy and allows for a cleaner isolation of the temporal evolution of the NS structure.

We present \textsc{troye} (Transitional Representation Of varYing Equation-of-state), a time-domain waveform model constructed to simulate this phenomenology. Using Bayesian inference, we quantify the detectability horizon for dynamic phase transitions in the era of Advanced LIGO and future detectors. Our approach is conceptually inspired by the IMR consistency test \cite{ghosh2016IMR}, but here we use the same cut-and-stitch idea to construct waveforms with an explicit, time-localized change in the tidal response.

\section{\label{sec:background}Background}
Unless stated otherwise, we use natural units ($c=G=1$) throughout this paper.

\subsection{Neutron Star Interior and Equation of State}
Neutron stars (NSs) provide a distinct environment for exploring the physical theories of dense matter, with an interior composed of three main layers: an outer crust, an inner crust, and a core \cite{page2006}. The crust is characterized by a lattice of nuclei embedded in a degenerate electron gas \cite{Lattimer2004}. In the inner crust, neutrons "drip" out of nuclei, forming a superfluid. At the transition to the core ($\rho \gtrsim 1.5 \times 10^{14}\, \text{g/cm}^3$), nuclei dissolve completely into a homogeneous fluid of neutrons, protons, electrons, and muons \cite{baym2018}. This canonical layering captures the consensus qualitative structure, while the precise transition densities and inner-core composition are EoS- and mass-dependent.

The properties of this ultra-dense matter are governed by the EoS, $p(\rho)$. The EoS determines the macroscopic structure of the star via the Tolman-Oppenheimer-Volkoff (TOV) equations for hydrostatic equilibrium in general relativity \cite{tov1939}:
\begin{equation}
    \begin{cases}
    \dfrac{dp}{dr} = -\dfrac{(p+\rho)(m + 4\pi r^3 p)}{r(r-2m)}, \\[10pt]
    \dfrac{dm}{dr} = 4\pi r^2 \rho.
    \end{cases}
\end{equation}
Integrating these equations from the center ($r=0, P=P_c$) to the surface ($p(R)=0$) for a given EoS yields the mass-radius relation, $M(R)$ \cite{ozel2016}. Since the microphysics of the inner core—potentially containing hyperons, pion condensates, or deconfined quark matter—is insufficiently understood \cite{huth2022, hyperon2017}, the EoS remains one of the largest uncertainties in nuclear astrophysics. This uncertainty is often parameterized by the "stiffness" of the EoS: a "stiff" EoS (high pressure) supports larger radii and higher maximum masses, while a "soft" EoS leads to more compact stars \cite{burgio2021}.

Current astrophysical constraints are derived from X-ray timing (e.g., NICER \cite{nicer2012, riley2019nicer}) and gravitational-wave (GW) observations \cite{abbott2018, de2018tidal}. The latter are particularly sensitive to the EoS through tidal interactions \cite{read2009}.

\subsection{Tidal Deformability and Phase Transitions}
In a binary system, the gravitational field of a companion induces a quadrupole moment $Q_{ij}$ in the NS. The response of the star to the external tidal field $\mathcal{E}_{ij}$ is quantified by the tidal deformability $\lambda$, defined as $Q_{ij} = -\lambda \mathcal{E}_{ij}$ \cite{hinderer2008, katerina2020}. The negative sign indicates that the induced quadrupole moment opposes the external tidal field, acting as a restorative force against the distortion.
To facilitate comparison between stars of different masses, we define the dimensionless tidal deformability $\Lambda$:
\begin{equation}
    \Lambda \equiv \frac{\lambda}{m^5} = \frac{2}{3} k_2 \left( \frac{R}{m} \right)^5 = \frac{2}{3} k_2 C^{-5},
\end{equation}
where $k_2$ is the tidal Love number (typically $0.05-0.15$ for realistic EoSs \cite{hinderer2008, poisson2014}) and $C = m/R$ is the stellar compactness. $\Lambda$ is highly sensitive to the stellar radius ($\Lambda \propto R^5$), making it an excellent probe of the EoS. Stiff EoSs typically yield $\Lambda \sim \mathcal{O}(10^3)$, while soft EoSs yield $\Lambda \sim \mathcal{O}(10^2)$ for a $1.4 M_\odot$ star \cite{abbott170817}.

The induced tidal deformation extracts energy from the orbit, accelerating the inspiral. This effect enters the GW phase evolution at 5th Post-Newtonian (5PN) order \cite{Flanagan2008}. Despite the high order, the large value of $\Lambda$ allows this effect to be measurable primarily in the late inspiral (at frequencies of a few hundred Hz, with the exact onset depending on the component masses and detector sensitivity) by advanced detectors \cite{read2009, katerina2020}.

A unique feature of the binary environment is the rapid increase in the strength and time-variation of the tidal field as merger approaches, which drives large internal stresses and can excite oscillation modes. In this regime, the relevant control parameters for microphysics (e.g., temperature, composition, chemical potentials, and stress anisotropies) may evolve quickly, and the star may cross a critical threshold for a change in its internal state. Motivated by the possibility of such rapid rearrangements, including strong hadron--quark transitions \cite{Drago2014, katerina2020phasetrans}, we consider the phenomenology of a time-dependent tidal response $\Lambda(t)$ rather than the static $\Lambda(m)$ assumed in standard LVK analyses. We also note that tidal interactions are not expected to generically increase the star's central density; analytic arguments show that tidal fields can instead stabilize compact stars against collapse, lowering the central density near marginal stability \cite{thorne1998, lai1996}.
If the transition is first-order, it may lead to an effective softening accompanied by radial contraction and a decrease in $\Lambda$. Alternatively, other structural changes could stiffen the effective response. In either case, a dynamical transition implies that the tidal imprint on the GW phase changes in time.

\subsection{NSBH Systems as a Laboratory}
We focus on NSBH systems as the primary laboratory for this search. In the parameter space relevant to ground-based detectors, NSs typically have masses of order $1$--$2\,M_\odot$, while the black-hole companion can range from a few to $\mathcal{O}(10)\,M_\odot$, leading to mass ratios $q=m_{\rm NS}/m_{\rm BH}\sim 0.05$--$0.3$. Such systems enter the LIGO band at $f\sim 20\,{\rm Hz}$ tens of seconds before merger (depending primarily on the chirp mass), while the characteristic merger frequency is typically in the few hundred Hz range and can vary substantially with the component masses. The detectable signal strength scales approximately as $\rho \propto 1/d_L$ (for fixed intrinsic parameters and orientation), so nearby NSBHs dominate the population of high-SNR events where subtle tidal physics is most accessible. Finally, whether the NS is tidally disrupted outside the horizon or instead plunges smoothly depends on the mass ratio, black-hole spin (especially if aligned), and NS compactness; disruption is favored for lower $m_{BH}$ and higher aligned $\chi_{BH}$, whereas higher $m_{BH}$ and low spin typically lead to direct plunge.

In BNS systems, the observable tidal parameter is the effective tidal deformability $\tilde{\Lambda}$, a mass-weighted combination of the individual NSs' $\Lambda_1$ and $\Lambda_2$ \cite{Flanagan2008, abbott170817}, defined as:
\begin{equation}
    \tilde{\Lambda} = \frac{16}{13} \frac{(m_1 + 12m_2)m_1^4 \Lambda_1 + (m_2 + 12m_1)m_2^4 \Lambda_2}{(m_1 + m_2)^5}.
\end{equation}
Disentangling the individual evolution of two stars, which might undergo transitions at different times, is degenerate.
In an NSBH system, assuming a "no-hair" black hole ($\Lambda_{BH}=0$), all tidal effects map to the single NS. This breaks the degeneracy, allowing for a cleaner isolation of the temporal evolution of the NS structure \cite{chen2020, kumar2017, cho2022measurability}. While the tidal signal in an NSBH event is generally weaker than in a BNS event due to the larger total mass, this is expected: higher-mass systems merge at lower frequencies and spend fewer cycles in the late inspiral where tidal phase corrections accumulate most strongly, and the effective tidal imprint is further suppressed at large mass ratios. The simplified systematics nevertheless make NSBHs an attractive target for searching for subtle dynamic effects.

To date, the LVK collaboration has reported a growing catalog of NSBH candidates. The first definitive detections, GW200105 and GW200115 \cite{abbott2021}, were consistent with plunging systems where the NS was swallowed whole, leaving no tidal imprint. Ambiguous events such as GW190814 \cite{abbott2020gw190814}, involving a $2.6 M_\odot$ secondary in the lower mass gap, highlight the challenge of distinguishing heavy NSs from light black holes. More recently, the O4 observing run yielded GW230529 \cite{abac2024gw230529, sanger2024tests}, another merger involving a compact object in the mass gap ($2.5 - 4.5 M_\odot$). In the O4a catalog, GW230518\_125908 is the only other candidate with component masses consistent with one object being a NS \cite{abac2025gwtc}. For both GW230529\_181500 and GW230518\_125908, LVK analyses with tidal waveform models report no meaningful constraint on the neutron-star tidal deformability, consistent with their SNRs and mass inferences \cite{abac2024gw230529, abac2025gwtc}. As detector sensitivity improves, the catalog of high-SNR NSBH events—potential candidates for observing tidal dynamics—is expected to expand significantly.

\section{\label{sec:method}Methodology}

To investigate the observability of equation-of-state phase transitions during the late inspiral, we employ a phenomenological approach that modifies the gravitational waveform to reflect a macroscopic change in the neutron star's structure. We introduce the \textsc{troye} waveform model, which simulates this dynamical effect by stitching together two distinct waveform approximants in the time domain, representing the binary evolution before and after the critical density threshold is crossed.

\subsection{Waveform Stitching}
The \textsc{troye} model is deliberately phenomenological. Its only working assumption is that the signal can be represented by two baseline tidal responses, characterized by $\Lambda_{pre}$ and $\Lambda_{post}$, with an effective change occurring at a time $t_{stitch}$. This parametrization is motivated by scenarios in which the NS internal structure changes rapidly during the late inspiral, producing an abrupt change in the effective tidal response. One example is a strong first-order phase transition once the core conditions cross a critical threshold, but the model does not rely on any specific trigger mechanism: any microphysics that yields a sudden change in the tidal response (e.g., anisotropic stresses, crystallization, or mode-driven rearrangement) would be captured by the same $(\Lambda_{pre},\Lambda_{post})$ description. We also note that tidal interactions are not expected to generically increase the star's central density; analytic arguments show that tidal fields can instead stabilize compact stars against collapse, lowering the central density near marginal stability \cite{thorne1998, lai1996}. Conceptually, this construction was inspired by the inspiral--merger--ringdown (IMR) consistency test \cite{ghosh2016IMR}, which checks GR by comparing information inferred from different portions of the signal. Our goal is different: we do not use the IMR-consistency infrastructure (e.g., splitting an observed signal into inspiral and post-inspiral for a consistency check), nor do we propose this as a replacement or extension of that test. Rather, we adopt the same general idea of separating waveform information in time/frequency, and here use it in a forward-modeling sense to represent an internal change in the NS.

The construction of a \textsc{troye} signal begins with the generation of two complete time-domain baselines, $h_{pre}(t)$ and $h_{post}(t)$, utilizing the \texttt{IMRPhenomNSBH} approximant \cite{thompson2020}. These waveforms share identical binary parameters (component masses, spins, luminosity distance, orientation, and sky location), differing solely in their tidal parameters. To avoid unphysical discontinuities in the gravitational-wave phase—which would dominate the matched-filter response and bias parameter estimation—we implement a dynamic phase alignment procedure. We compute the instantaneous phase of both polarizations at the stitching epoch, deriving a phase offset $\Delta \phi = \phi_{pre}(t_{stitch}) - \phi_{post}(t_{stitch})$. The post-transition waveform is then rotated in the complex plane by a factor $e^{i\Delta \phi}$. While precessing systems would require a simultaneous rotation of the orbital plane to maintain continuity, this simple phase rotation is sufficient for the non-precessing (aligned-spin) systems considered in this work.

The transition between the two physical states is not modeled as instantaneous; rather, we define a finite temporal ``blending window'' of duration $\tau$ centered at $t_{stitch}$. While the microscopic nucleation timescales could be sub-millisecond \cite{Prasad2018}, the macroscopic duration of the phase transition is governed by the propagation speed of the conversion front across the stellar interior. We adopt a phenomenological duration of $\tau = 10$ ms for this study. In the late inspiral frequency band, this duration spans several orbital cycles ($T_{orb} \sim 2-3$ ms), representing a regime where the structural change is dynamically significant but not effectively instantaneous. This choice serves as a conservative benchmark, testing the recoverability of a diffusive transition while ensuring numerical smoothness. Within this window, the waveforms are superimposed using a transition function $w(t)$. While the architecture supports various taper choices (e.g., Tukey, Hamming), we adopt a raised-cosine (Hann) window because it is smooth at the endpoints ($w$ and its first derivative vanish at $t_\pm$), which suppresses edge artifacts and spurious broadband power that would otherwise contaminate matched filtering. Since the phase is explicitly aligned at $t_{stitch}$, the stitched waveform is expected to depend only weakly on the detailed taper shape, provided the window is smooth and confined to the same duration $\tau$; the raised-cosine therefore provides a simple, numerically stable default.
The final stitched strain $h_{stitched}(t)$ is given by:

\begin{equation}
h_{stitched}(t) = 
\begin{cases} 
h_{pre}(t), & t < t_{-} \\ \\
\begin{aligned}
&[1 - w(t)] h_{pre}(t) \\
&+ w(t) h_{post}^{aligned}(t),
\end{aligned} & t_{-} \leq t \leq t_{+} \\ \\
h_{post}^{aligned}(t), & t > t_{+}
\end{cases}
\end{equation}
where $t_{\pm} \equiv t_{stitch} \pm \tau/2$.

\begin{figure*}[t]
    \centering
    \includegraphics[width=1\linewidth]{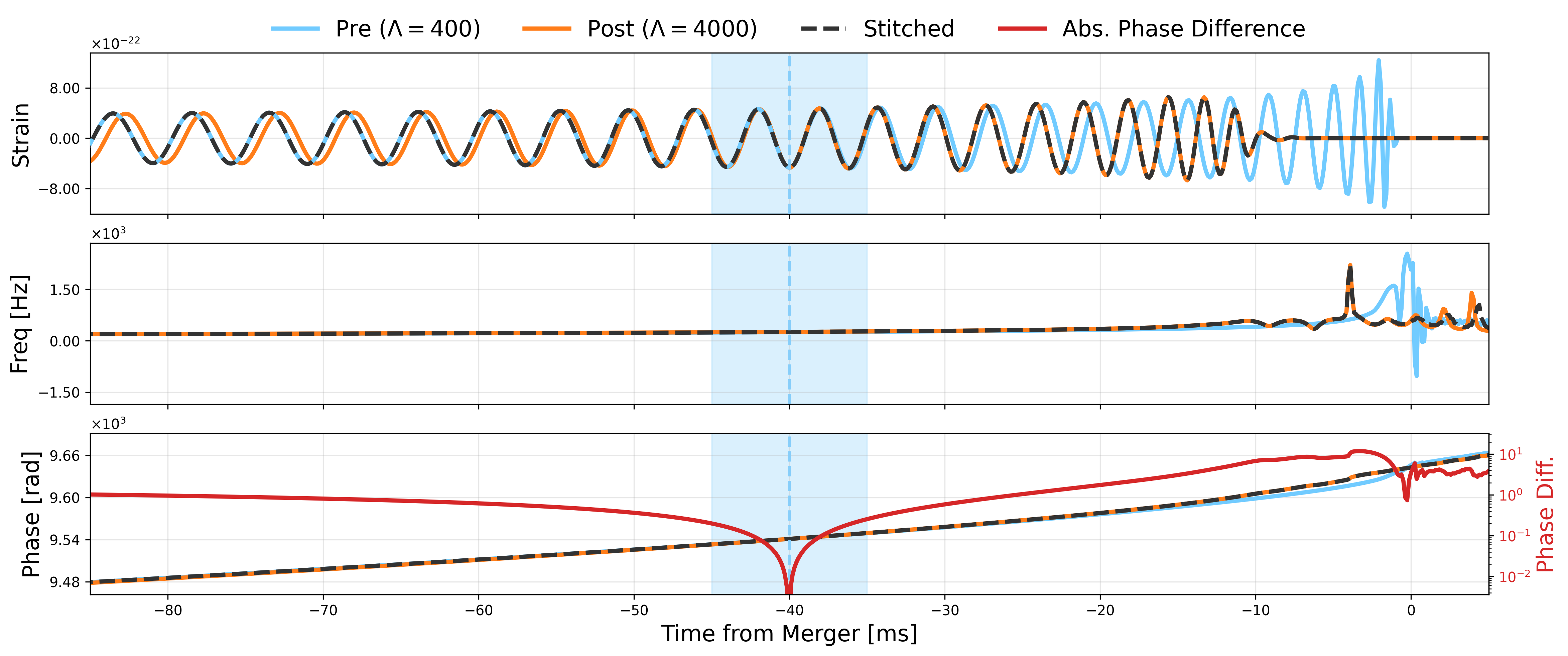}
    \caption{
    Detailed diagnostic of the \textsc{troye} stitching and phase alignment methodology. The panels display the evolution of the GW signal for a pre-transition model ($\Lambda=400$, solid light blue line), a post-transition model ($\Lambda=4000$, solid orange line), and the resulting stitched waveform (dashed black line). Top Panel: The strain amplitude $h(t)$. Middle Panel: The instantaneous frequency evolution $f(t)$. Bottom Panel: The unwrapped instantaneous phase (left axis) and the absolute phase difference between the aligned pre- and post-transition waveforms (red curve, right logarithmic axis). The vertical dashed line marks the stitching time $t_{stitch}$. Note that the dynamic phase alignment ensures the relative phase difference vanishes ($|\Delta\phi| \to 0$) precisely at $t_{stitch}$, guaranteeing a smooth ``hand-off'' between the models without discontinuities. A measurable phase discrepancy accumulates only as the system evolves away from $t_{stitch}$ in either direction, encoding the distinct tidal signatures of the two physical states. Note how the stitched signal perfectly tracks the pre-transition evolution (hiding the blue trace) prior to the window (light blue shaded region) and seamlessly transitions to follow the post-transition evolution (hiding the orange trace) thereafter, ensuring phase continuity and a physically smooth transition in the tidal parameter.}
    \label{fig:stitching}
\end{figure*}

It is important to note that this model captures the primary kinematic signature of the transition—the macroscopic change in tidal deformability—by stitching vacuum solutions; it does not self-consistently solve the Einstein-hydrodynamic equations for a fluid undergoing latent heat release or non-adiabatic oscillations. Fig.~\ref{fig:stitching} illustrates the phase-alignment and blending procedure and the resulting phase continuity at $t_{stitch}$.

\subsection{Waveform Injection Campaign}
\label{subsec:injection_campaign}

The injection campaign was designed to capture the phenomenological signature of the phase transition within a realistic parameter space. The total duration of the analyzed waveform segments was set to $128\,\mathrm{s}$ to ensure the capture of the full inspiral. In practice, source waveforms were generated using the \texttt{IMRPhenomNSBH} approximant with signal durations ranging from $60\,\mathrm{s}$ to $100\,\mathrm{s}$, which were then injected into $128\,\mathrm{s}$ of colored noise (see Sec.~\ref{subsec:detector_response}). The model stitch time $t_{stitch}$ was consistently fixed to $40\,\mathrm{ms}$ prior to the merger time $t_c$ (defined by \texttt{IMRPhenomNSBH} as the peak strain amplitude).

\begin{figure}[b]
    \centering
    \includegraphics[width=1\linewidth]{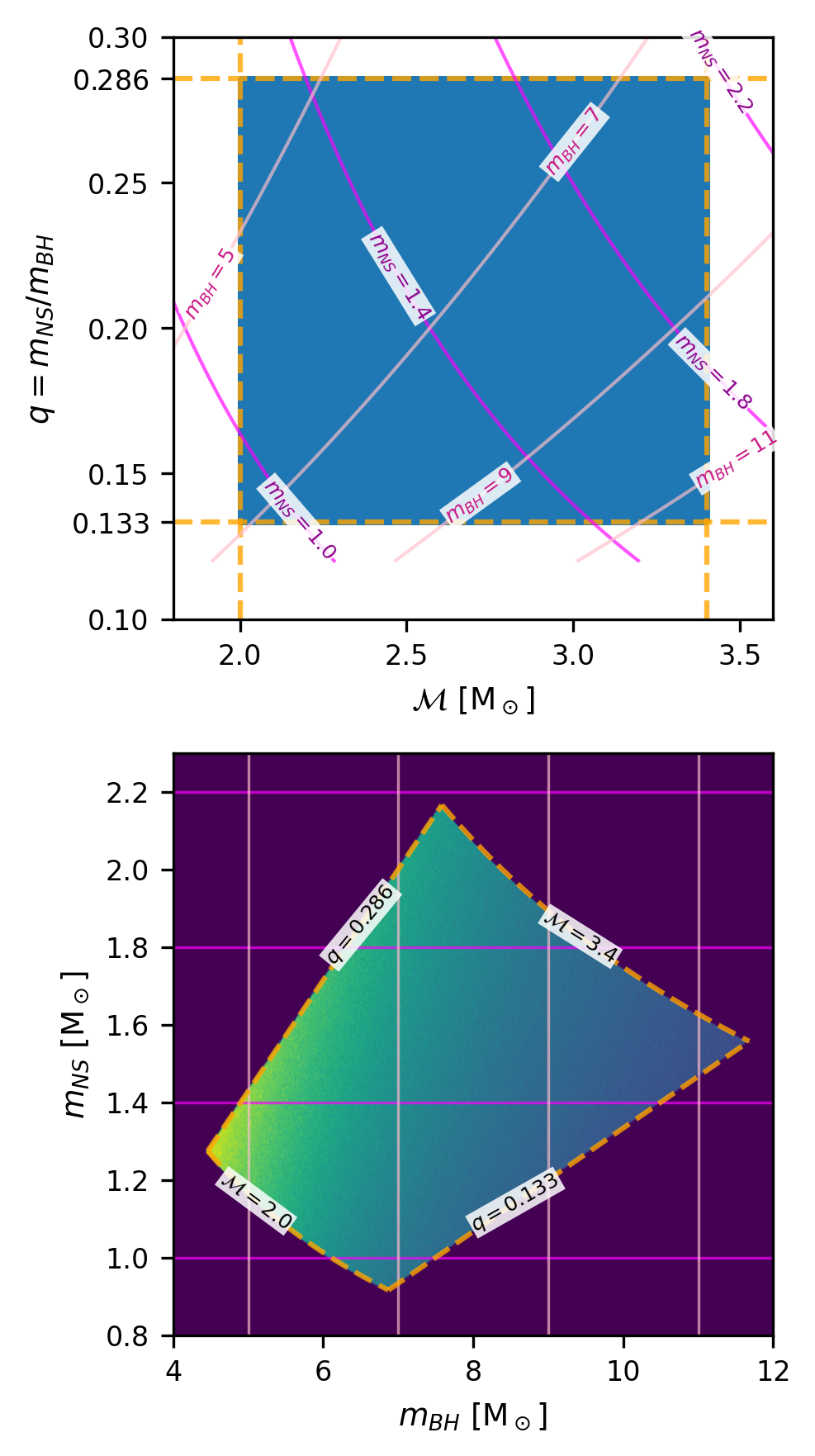}
    \caption{The chosen mass prior distribution of the generated waveforms.
    On the top panel, a uniform distribution in the chirp mass ($\mathcal{M}$) - mass ratio ($q$) space.
    On the bottom, a mapping of that same prior to the $m_{BH}-m_{NS}$ space.
    The color brightness represents the density of the Jacobian for the transformation.
    The solid pink (magenta) lines show the inverse transformation of equal $m_{BH}$ ($m_{NS}$) lines, respectively, to the $\mathcal{M}-q$ plane. The dashed orange lines show the prior "bounding box" on both planes.}
    \label{fig:mass_space}
\end{figure}

\begin{figure}[b]
    \centering
    \includegraphics[width=1\linewidth]{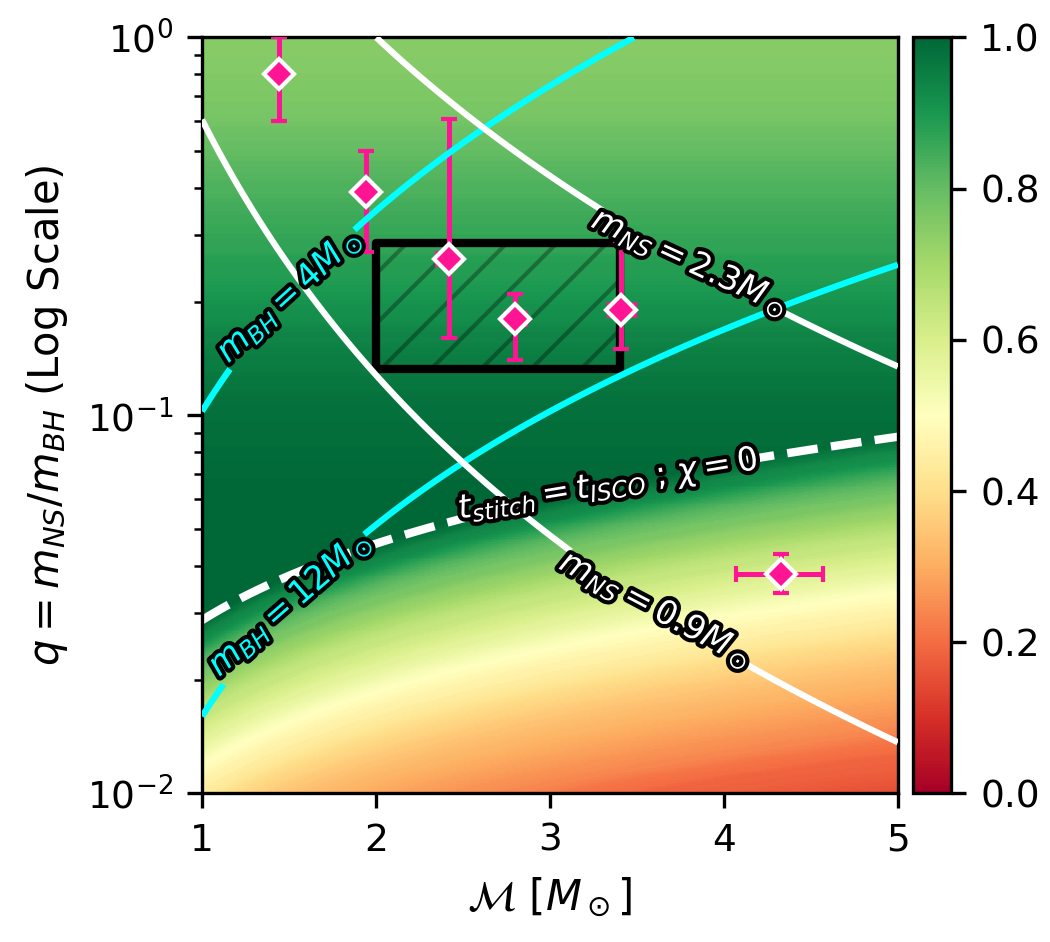}
    \caption{
    Domain-validity map for the injection campaign in the $\mathcal{M}$--$q$ plane. Colors show the probability of a safe inspiral (marginalized over other intrinsic parameters), estimated from $N=1000$ random draws per pixel with $\chi_{BH}\sim U(0,0.99)$ and $\Lambda_{NS}\sim U(10,4900)$. A draw is labeled safe if it (i) avoids early plunge relative to the stitching time ($t_{\rm ISCO}<t_{\rm stitch}=40\,\mathrm{ms}$) and (ii) does not tidally disrupt ($M_{\rm rem}=0$ from the Foucart criterion \cite{foucart2012}, with $C$ inferred from $\Lambda$ via the Yagi--Yunes universal relations \cite{YagiYunes2017}). Solid white (cyan) contours denote constant $m_{NS}\in\{0.9,2.3\}\,M_\odot$ ($m_{BH}\in\{4,12\}\,M_\odot$). The dashed white curve is the conservative zero-spin plunge boundary ($\chi_{BH}=0$), above which early plunge is excluded for any spin. Magenta diamonds mark LVK events \cite{abbott190425, abac2024gw230529, abbott2021, gwtc3, abac2025gwtc}; from left to right: GW190425, GW230529, GW200115, GW230518, GW200105, GW191219. The hatched box shows our adopted prior, $\mathcal{M}\in[2.0,3.4]\,M_\odot$ and $q\in[0.133,0.286]$, chosen to maximize the yield of valid signals while covering realistic NSBH systems.
    }
    \label{fig:sweet_spot}
\end{figure}

We selected uniform prior distributions for the chirp mass $\mathcal{M} \in [2.0, 3.4]\,M_\odot$ and mass ratio $q \in [0.133, 0.286]$. As illustrated in Fig.~\ref{fig:mass_space}, this choice maps to a specific region in the component mass plane. This prior range is motivated by the domain validity analysis shown in Fig.~\ref{fig:sweet_spot}: the mass ratio is chosen high enough to ensure $t_{stitch}$ occurs before the Innermost Stable Circular Orbit ($t_{stitch} < t_{ISCO}$, where $t_{ISCO}$ corresponds to the time when the binary separation reaches the dimensionless ISCO radius $R_{ISCO}/M_{BH}$), thereby avoiding an early plunge, yet low enough to minimize the probability of a Tidal Disruption Event (TDE). The chirp mass range ensures the inclusion of realistic neutron stars ($m_{NS} < 2.35\,M_\odot$), predominantly with $m_{NS} > 1\,M_\odot$, while allowing for a small parameter region down to $m_{NS} \approx 0.9\,M_\odot$.

For the bulk of the simulation campaign (comprising 100 recovered events), we adopted a strategy to maximize computational efficiency. The primary objective was to determine the recoverability of the tidal shift $\Delta \Lambda$ rather than to constrain the full 15-dimensional parameter space of the binary. Therefore, we fixed the majority of the source parameters to their injected values using Dirac delta function priors during the recovery. This includes the intrinsic parameters (mass ratio $q$ and black hole spin $\chi_{BH}$) as well as all extrinsic parameters (sky location $\alpha, \delta$, luminosity distance $d_L$, and inclination $\iota$). While these parameters were drawn from the broad distributions detailed in Table~\ref{tab:priors} for the injection, fixing them during recovery allows the sampler to focus its resolution specifically on the chirp mass $\mathcal{M}$ and the pre- and post-transition tidal deformabilities, thereby avoiding the computational expense of exploring nuisance parameter degeneracies. To ensure the robustness of this approximation, a subset of ``stress test'' analyses was performed with expanded priors (see Sec.~\ref{sec:stress_tests}), confirming that the recoverability of the phase transition persists even when marginalized over uncertainties in the other parameters. Sampler settings were tuned for throughput, with the number of live points (\texttt{nlive}) set to 100.

\subsection{Detector Response and Noise Model}
\label{subsec:detector_response}

To ensure geometric consistency across a network of interferometers, the stitching procedure is applied directly to the gravitational-wave polarizations, $h_+(t)$ and $h_\times(t)$, in the source frame. This guarantees that the phase transition event arrives at each detector with the correct relative timing and amplitude corrections derived from the antenna pattern functions $F_+$ and $F_\times$. The detector strain $h_i(t)$ is computed via standard projection \cite{thorne1987gravitational}:
\begin{equation}
\begin{split}
h_i(t) = & \, F_{+,i}(\alpha, \delta, \psi) h_+(t - \Delta t_i) \\
         & + F_{\times,i}(\alpha, \delta, \psi) h_\times(t - \Delta t_i),
\end{split}
\end{equation}
where $\Delta t_i$ is the time delay relative to the geocenter.

Simulated data were generated by injecting these waveforms into Gaussian noise colored by a specific Power Spectral Density (PSD). We utilized the \texttt{lalsimulation} implementation of the Advanced LIGO ``Zero Detuning, High Power'' (ZDHP) design sensitivity \cite{abbott2020prospects}. This configuration corresponds to a signal-recycling cavity tuned for broadband sensitivity with high circulating laser power, serving as a standard stationary noise reference for parameter estimation studies. Consequently, the results presented here should be interpreted as an optimistic (stationary-noise) baseline; in real interferometer data, non-Gaussian transients overlapping the final inspiral can bias evidence and broaden posteriors, motivating glitch-robust studies described in Sec.~\ref{subsec:limitations}.

To ensure that our analysis targets the tidal evolution during the inspiral rather than the complex hydrodynamics of disruption, we performed a post-hoc validity check on the simulation catalog. We utilized the semi-analytical Foucart criterion \cite{foucart2012} to predict the remnant disk mass ($M_{rem}$) for each injected event. A system is classified as a Tidal Disruption Event (TDE) if $M_{rem} > 0$. Because the Foucart model depends on the neutron-star compactness $C=m/R$, we convert each injected $\Lambda_{NS}$ to $C$ using the Yagi--Yunes universal Love--compactness relations \cite{YagiYunes2017}. We found that our injection priors successfully favored non-disruptive mergers; only 6 out of the 100 simulated events were classified as likely TDEs. Notably, the phase transition signature remained robustly detectable in five of these six marginal cases, suggesting that the late-inspiral tidal footprint persists even in systems approaching the disruption limit.

\subsection{Bayesian Inference Framework}
\label{subsec:bayesian_framework}

To infer the properties of the source system from noisy detector data, we employ Bayesian inference \cite{ThraneTalbot2019}. We compute the posterior probability density of the source parameters $\vec{\theta}$ given the data $d$:
\begin{equation}
    p(\vec{\theta}|d) = \frac{\mathcal{L}(d|\vec{\theta}) \pi(\vec{\theta})}{\mathcal{Z}(d)},
\end{equation}
where $\mathcal{L}(d|\vec{\theta})$ is the likelihood assuming Gaussian noise \cite{Finn1992}, $\pi(\vec{\theta})$ is the prior probability distribution, and $\mathcal{Z}(d)$ is the Bayesian evidence.

We utilize the \texttt{bilby} inference library \cite{bilby2019}, which wraps the \textsc{troye} model within a \texttt{WaveformGenerator} to handle domain conversion and frequency-domain likelihood evaluation. To sample the posterior, we employed the \texttt{dynesty} dynamic nested sampling algorithm \cite{Speagle2020}. Unlike standard MCMC methods, nested sampling is designed to efficiently compute the evidence $\mathcal{Z}$. This is essential for our analysis, as we rely on the Bayes Factor ($\mathcal{B} = \mathcal{Z}_{\textsc{troye}} / \mathcal{Z}_{null}$) to quantify the statistical support for the dynamic phase transition model against the static null hypothesis \cite{katerina2020phasetrans}.

\begin{table}[t]
    \centering
    \caption{\label{tab:priors} Prior distributions used for the injection and recovery of the simulated NSBH population. The mass ratio is defined as $q = m_{NS}/m_{BH} \le 1$. Extrinsic parameters (sky location, orientation) were drawn from isotropic distributions. The chosen mass ranges correspond to a light Black Hole and a heavy Neutron Star, targeting the "lower mass gap" region ($2.5-4.5 M_\odot$) where tidal interactions are most likely to be observable in NSBH mergers.}
    \begin{ruledtabular}
    \begin{tabular}{l c c c l}
    \textbf{Parameter} & \textbf{Prior} & \textbf{Min} & \textbf{Max} & \textbf{Description} \\
    \hline
    $\mathcal{M}$ ($M_{\odot}$) & Uniform & $2.00$ & $3.40$ & Chirp Mass \\
    $q$ & Uniform & $0.133$ & $0.286$ & Mass Ratio \\
    $\chi_{BH}$ & Uniform & $0.0$ & $0.99$ & BH Spin Magnitude \\
    $\chi_{NS}$ & Delta & $0.0$ & $0.0$ & NS Spin Magnitude \\
    $\Lambda_{pre}$ & Uniform & $10$ & $4900$ & Initial Tidal Deform. \\
    $\Lambda_{post}$ & Uniform & $10$ & $4900$ & Final Tidal Deform. \\
    \hline
    $d_L$ (Mpc) & Comoving & $10$ & $100$ & Luminosity Distance \\
    $\alpha$ & Uniform & $0$ & $2\pi$ & Right Ascension \\
    $\delta$ & Cosine & $-\pi/2$ & $\pi/2$ & Declination \\
    $\iota$ & Sine & $0$ & $\pi$ & Inclination Angle \\
    \end{tabular}
    \end{ruledtabular}
\end{table}

\section{Results} \label{sec:results}
We evaluate the recoverability of dynamic phase transitions by analyzing an injection campaign of 100 simulated NSBH signals. 

\subsection{Detection Criteria}
To systematically classify the simulation results, we introduce two boolean indicators based on the properties of the recovered posterior distributions.

\begin{enumerate}
    \item \textbf{SP? (Separated Posteriors):} This flag indicates whether the sampler successfully resolved distinct values for the pre- and post-transition tidal deformabilities. It is marked \textbf{True} if the 90\% credible intervals of the recovered distributions for $\Lambda_{pre}$ and $\Lambda_{post}$ do not overlap.

    \item \textbf{PT? (Phase Transition Detected):} This is the final decision metric for the study. A phase transition is considered definitively detected only if the signal provides strong statistical evidence \textit{and} the inferred parameters show a significant shift distinct from the measurement uncertainty. The condition is defined as:
    \begin{equation}
        PT? \equiv \ln \mathcal{B} > 5.0 \land |\Delta \bar{\Lambda}| > \max(\sigma_{pre}, \sigma_{post}),
    \end{equation}
    where $\ln \mathcal{B}$ is the natural log Bayes Factor favoring the phase-transition hypothesis ($\mathcal{H}_{\textsc{troye}}$) over the null hypothesis ($\mathcal{H}_{0}$), corresponding to ``strong'' evidence on the Jeffreys scale \cite{jeffreys1998theory}. The second term ensures that the magnitude of the shift in posterior means, $|\Delta \bar{\Lambda}| \equiv |\bar{\Lambda}_{post} - \bar{\Lambda}_{pre}|$, exceeds the standard deviation ($\sigma$) of the widest individual posterior. Note that when we discuss detectability thresholds (e.g., $|\Delta\Lambda| \gtrsim 400$) in the context of the injection campaign results, we refer to the true \textit{injected} magnitude. However, for an individual detection candidate (flagged as \textbf{PT?}), we rely strictly on the recovered posterior means.
\end{enumerate}

\begin{figure}
    \centering
    \includegraphics[width=1\linewidth]{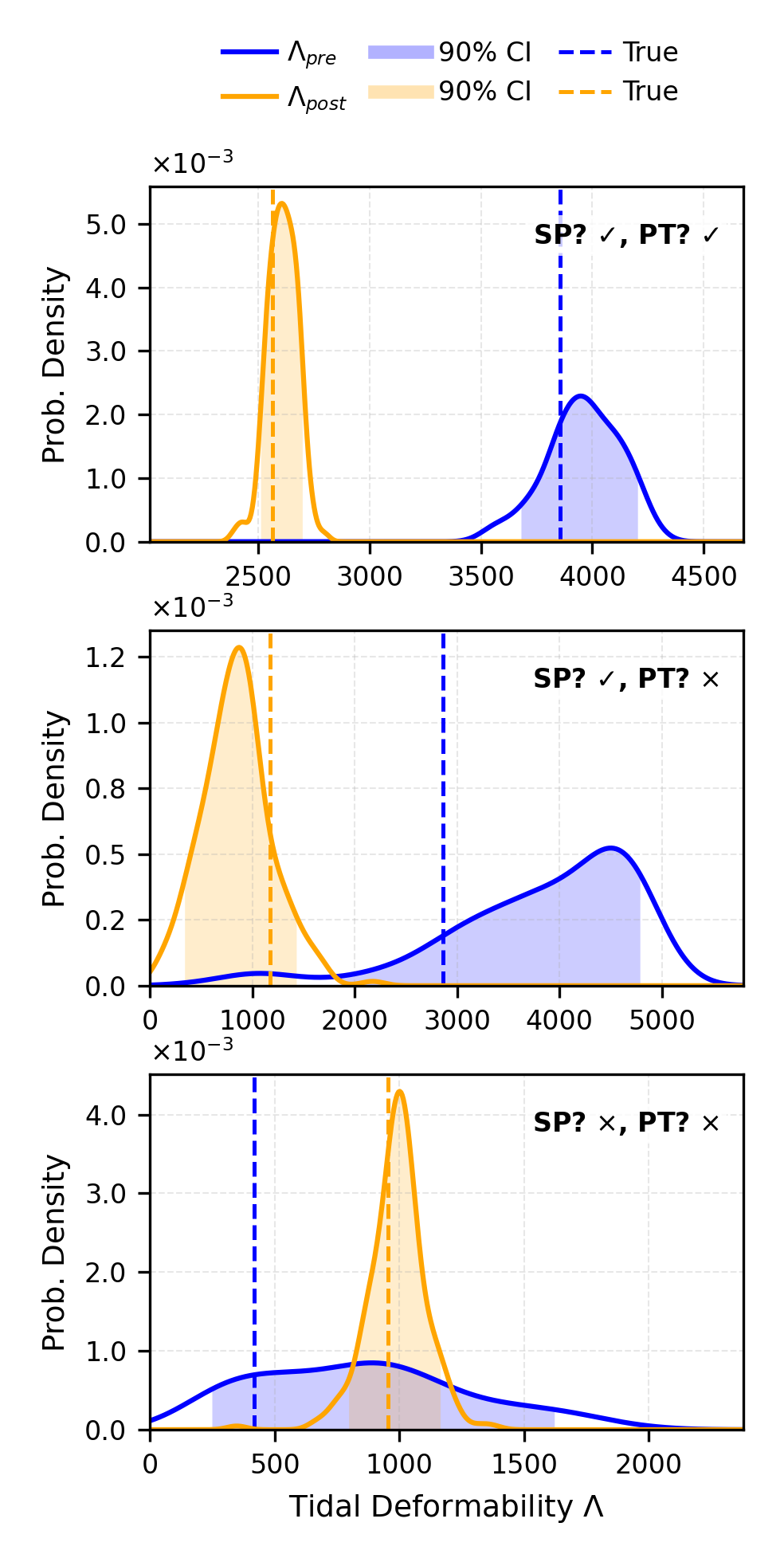}
    \caption{Representative posterior distributions for $\Lambda_{pre}$ (blue) and $\Lambda_{post}$ (orange) illustrating three distinct detection scenarios. Dashed vertical lines indicate the true injected values, while shaded regions denote 90\% credible intervals (CIs). 
    \textbf{Top panel:} A successful detection (SP $\checkmark$, PT $\checkmark$). The distributions are clearly distinct with non-overlapping CIs and strong Bayesian evidence ($\ln \mathcal{B} > 5.0$). 
    \textbf{Middle panel:} A visually separated case with insufficient evidence (SP $\checkmark$, PT $\times$). While the CIs do not overlap, the broad uncertainty in the pre-transition posterior results in a Bayes factor below the threshold. 
    \textbf{Bottom panel:} A clear non-detection (SP $\times$, PT $\times$) where the posteriors heavily overlap.}
    \label{fig:detection_scenarios}
\end{figure}

\begin{figure}[b]
    \centering
    \includegraphics[width=1\linewidth]{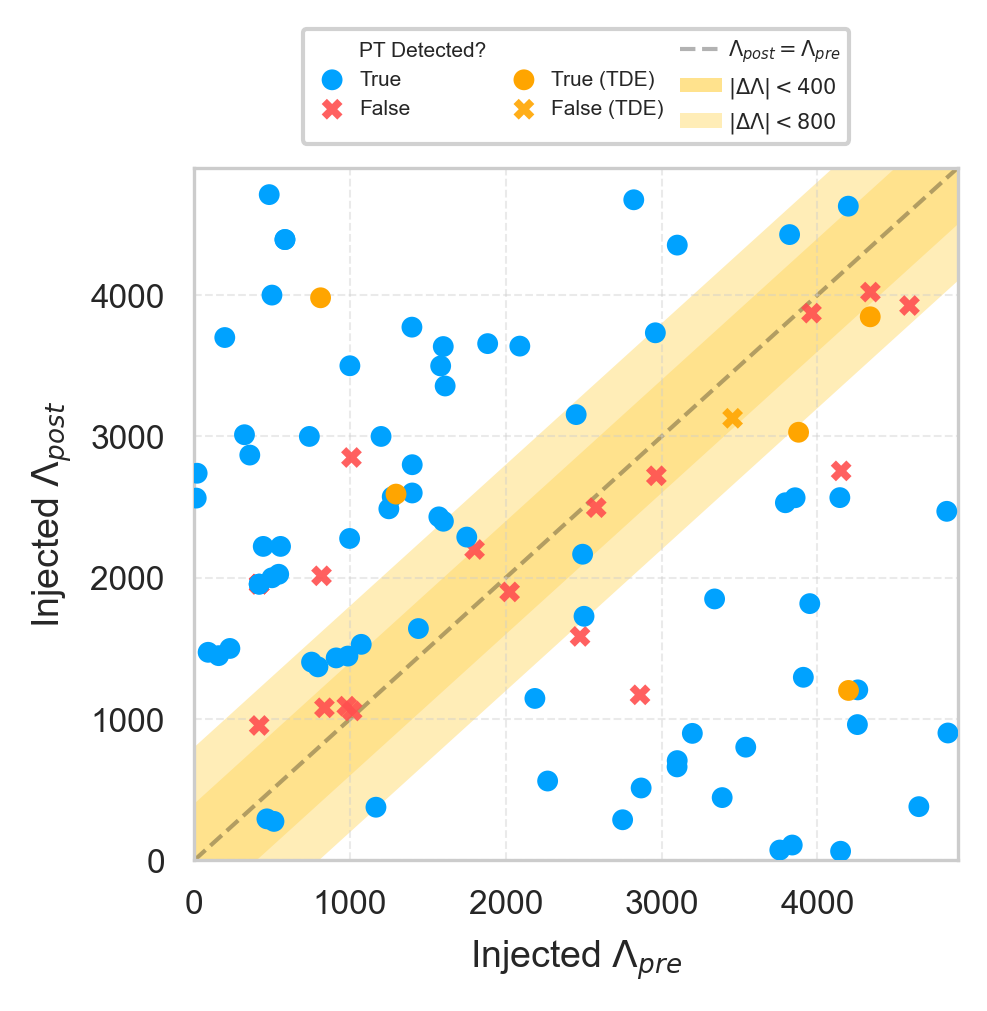}
    \caption{Distribution of simulated events in the $\Lambda_{pre}$ vs. $\Lambda_{post}$ plane, color-coded by detection status. Blue circles indicate successful phase transition detections (PT?=True), while red crosses denote non-detections. Orange circles (crosses) indicate a successful (unsuccessful) detection that was classified as a TDE by the Foucart criterion \cite{foucart2012}, with $C$ inferred from $\Lambda$ via the Yagi--Yunes universal relations \cite{YagiYunes2017}. The dashed grey diagonal line represents the null hypothesis ($\Lambda_{pre} = \Lambda_{post}$). The shaded regions highlight zones of subtle transitions, with $|\Delta \Lambda| < 400$ (darker yellow) and $|\Delta \Lambda| < 800$ (lighter yellow). The clustering of non-detections within the narrow band around the diagonal demonstrates that a minimum shift magnitude is required for the sampler to resolve the transition against the detector noise.}
    \label{fig:lambda_vs_lambda}
\end{figure}

Representative examples of these detection outcomes are shown in Fig.~\ref{fig:detection_scenarios}. Events failing the \textbf{PT?} condition (e.g., separated posteriors but insufficient evidence, or high evidence with overlapping error bars) are classified as non-detections. Using this criterion, we find that transitions with $|\Delta\Lambda| \gtrsim 400$ are consistently recoverable at $d_L \le 100$ Mpc. Fig.~\ref{fig:lambda_vs_lambda} provides an all-events summary in the $(\Lambda_{pre},\Lambda_{post})$ plane, highlighting the approximate $|\Delta\Lambda|\sim 400$ detectability boundary at which the transition becomes visible for nearby sources.

\subsection{SNR Gain and Phase Coherence}
\label{subsec:snr_gain_analysis}

\begin{figure}[b]
    \centering
    \includegraphics[width=\columnwidth]{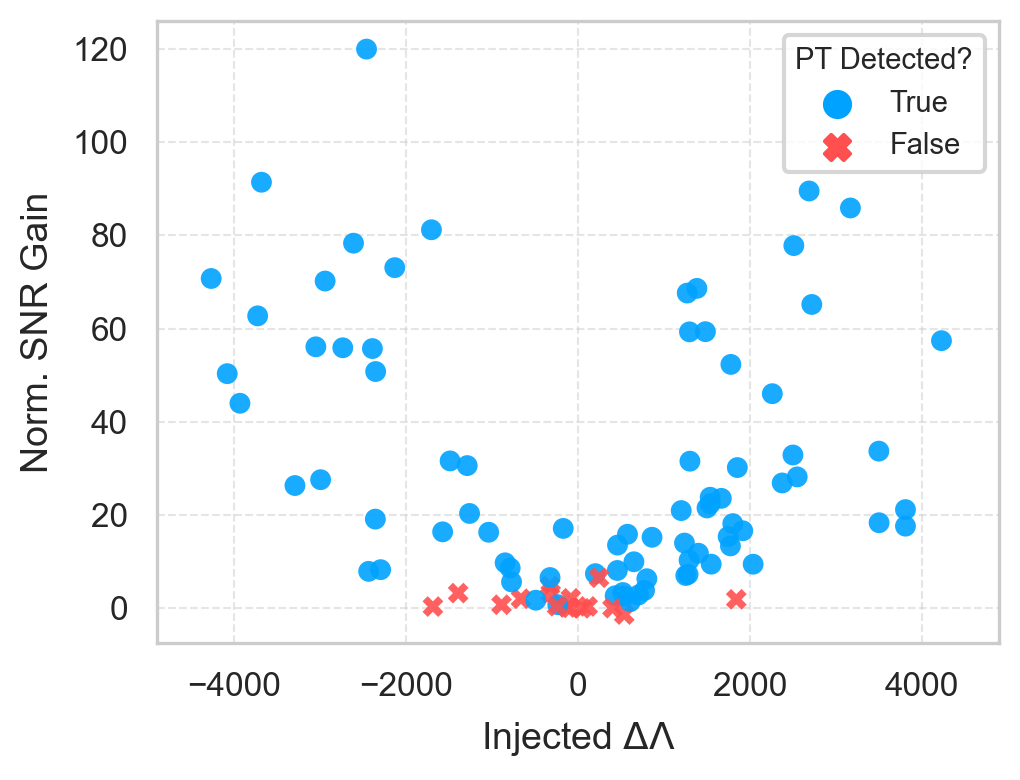}
    \caption{\label{fig:snr_scaling_signed} Normalized SNR gain ($\Delta\rho \cdot d_L \cdot \mathcal{M}^{-5/6}$) as a function of the injected tidal deformability shift $\Delta\Lambda$. The Y-axis represents the \textit{differential} contribution of the \textsc{troye} model. To address the scale comparison: the total normalized $\text{SNR}_{network}$ of the inspiral signal (the baseline) is approximately $2000$ in these units. Thus, observed gains of $\sim 50-100$ correspond to a $\sim 2-5\%$ improvement over the base signal. The distribution did not exhibit a statistically significant asymmetry in the SNR Gain of softening and stiffening events.}
\end{figure}

\begin{figure}[b]
    \centering
    \includegraphics[width=\columnwidth]{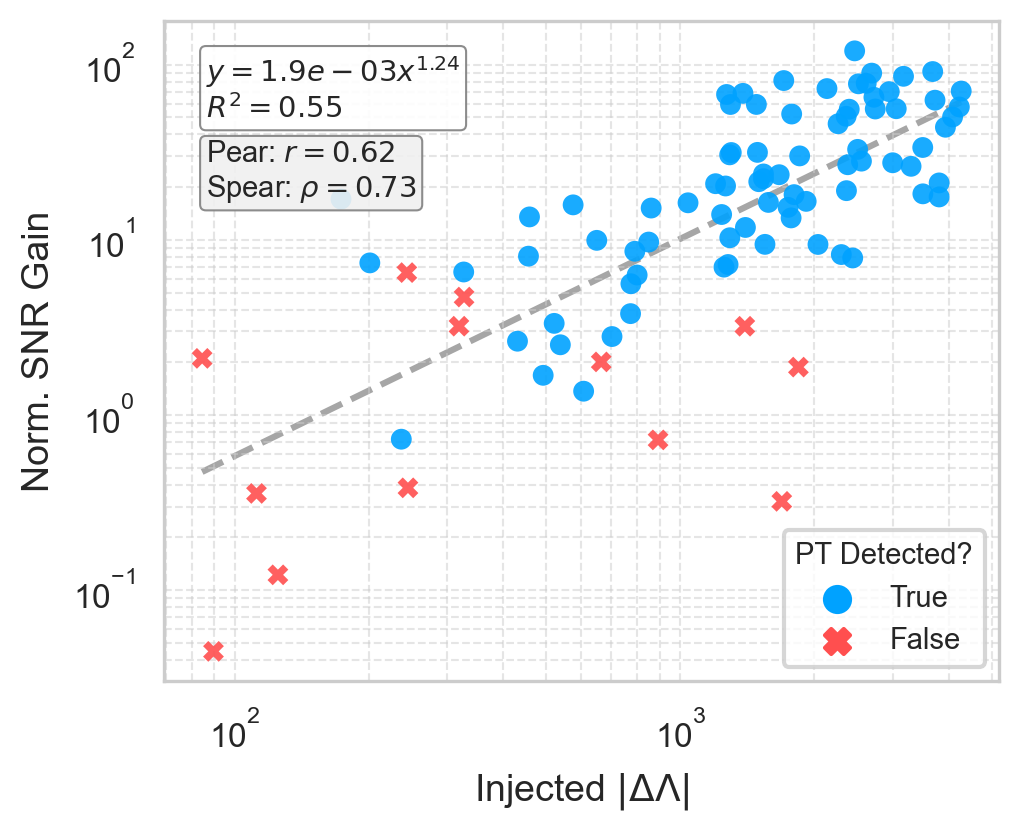}
    \caption{\label{fig:snr_scaling_log} Log-log projection of the SNR gain versus the absolute transition magnitude $|\Delta\Lambda|$. The data exhibits an approximate power-law trend scaling as $\sim |\Delta\Lambda|^{1.24}$. The significant scatter ($R^2 \approx 0.55$) reflects the dependence of the gain on the specific binary parameters which vary across the population.}
\end{figure}

The detectability of a tidal transition is fundamentally determined by the accumulated phase mismatch between the true ``transitional'' signal (\textsc{troye}) and the standard ``null hypothesis'' (a constant-$\Lambda$ waveform). Here, we derive the theoretical scaling of the SNR gain and compare it with the simulation results.

\subsubsection{Derivation of SNR Gain}

To quantify the advantage of the dynamic model over standard approximants, we consider the Matched Filter SNR. Let the true GW signal be $h_{tr}(f)$ and the null template be $h_{0}(f)$. We assume the transition affects the phase evolution $\Psi(f)$ significantly more than the amplitude $A(f)$. This approximation is justified because the amplitude envelope is dominated by the chirp mass $\mathcal{M}_c$ and luminosity distance $d_L$, which are identical for the two models, whereas tidal effects manifest primarily as cumulative phase corrections \cite{Flanagan2008}. Thus, the frequency-domain waveforms can be approximated as:
\begin{equation}
    \tilde{h}_{tr}(f) \approx A(f) e^{i\Psi_{tr}(f)}, \quad \tilde{h}_{0}(f) \approx A(f) e^{i\Psi_{0}(f)}
\end{equation}
The optimal SNR available in the signal, $\rho_{opt}$, is defined by the inner product of the signal with itself \cite{flanagan1998measuring, thorne1987gravitational}:
\begin{equation}
    \rho_{opt} = \sqrt{(h_{tr} | h_{tr})} \approx \sqrt{4 \int_{0}^{\infty} \frac{A^2(f)}{S_n(f)} df}
\end{equation}
where $S_n(f)$ is the one-sided Power Spectral Density (PSD) of the detector. However, if this signal is recovered using the incorrect null template $h_0$, the recovered SNR $\rho_{rec}$ is the projection of the signal onto that template:
\begin{equation}
    \rho_{rec} = \frac{(h_{tr} | h_{0})}{\sqrt{(h_{0} | h_{0})}} = \frac{4 \Re \int_{0}^{\infty} \frac{A^2(f) e^{i[\Psi_{tr}(f) - \Psi_{0}(f)]}}{S_n(f)} df}{\rho_{opt}},
\end{equation}
where we have used $\sqrt{(h_0|h_0)} \approx \sqrt{(h_{tr}|h_{tr})} = \rho_{opt}$, consistent with our assumption that the amplitude envelopes are identical.
Defining the phase discrepancy as $\Delta \Phi(f) = \Psi_{tr}(f) - \Psi_{0}(f)$, the numerator simplifies to a weighted integral over the cosine of the phase error. The \textbf{SNR Gain} $\Delta \rho$ achieved by using the correct model is the difference between the optimal and recovered SNRs:
\begin{equation}
    \Delta \rho = \rho_{opt} - \rho_{rec} \propto \int_{f_{trans}}^{f_{cut}} \frac{A^2(f)}{S_n(f)} \underbrace{\left[ 1 - \cos\big(\Delta \Phi(f)\big) \right]}_{\text{Coherence Loss Term}} df.
\end{equation}

This derivation highlights two regimes:
\begin{enumerate}
    \item \textbf{Small deviations:}
    When the phase difference is small ($\Delta\Phi \ll 1$), we expand
    $1-\cos(\Delta\Phi)\simeq \tfrac12(\Delta\Phi)^2$.
    Since the leading tidal contribution to the inspiral phase is linear in the tidal parameter, the model-to-model phase difference scales as
    $\Delta\Phi \propto \Delta\Lambda$ \cite{Flanagan2008}, and therefore
    \begin{equation}
        \Delta\rho \propto \int_{f_{\rm trans}}^{f_{\rm cut}} \frac{A^2(f)}{S_n(f)}(\Delta\Lambda)^2\,df
        \propto (\Delta\Lambda)^2,
    \end{equation}
    as long as the integration domain is effectively unchanged.
    We do not assume $A(f)$ is constant in frequency; for fixed intrinsic parameters and distance, $A^2(f)/S_n(f)$ is independent of $\Delta\Lambda$ and acts as a fixed weighting kernel. However, in our model, the limits of the integral are not fixed: the transition location and especially the termination of the signal can shift with $\Delta\Lambda$, modifying the overall coefficient.

    A rough bound for the Taylor regime follows from the accumulated tidal phase by merger. For a canonical NSBH system (e.g., $1.4M_\odot+5M_\odot$), a deformability of $\Lambda\approx 400$ produces an order-$2\pi$ tidal phase shift relative to the zero-tides case by merger. Scaling linearly, requiring $\Delta\Phi\lesssim 1$ rad implies
    $|\Delta\Lambda|\lesssim 400/(2\pi)\approx 60$.
    Since detectable transitions are typically $|\Delta\Lambda|\sim 10^2$--$10^3$ (See Fig.~\ref{fig:lambda_vs_lambda}), this quadratic regime lies largely below the detection horizon and is unlikely to be cleanly resolved in our SNR-gain analysis.

    \item \textbf{Large deviations:}
    For larger shifts, $\Delta\Phi$ becomes $\gtrsim 1$ over an increasing portion of the post-transition band, so $\cos(\Delta\Phi)$ oscillates rapidly and averages out; the mismatch term behaves as $1-\cos(\Delta\Phi)\sim \mathcal{O}(1)$ wherever the phase has decorrelated. In this regime, the growth of $\Delta\rho$ is controlled by two coupled effects:
    (i) an increasing fraction of the weighted bandwidth becomes effectively incoherent, and
    (ii) the available bandwidth itself changes because the signal duration (and thus the effective cutoff frequency) depends on $\Delta\Lambda$.
    In particular, $\Delta\Lambda>0$ tends to shorten the signal (earlier termination relative to the no-transition or constant-$\Lambda$ case), reducing the upper integration limit and suppressing the gain, while $\Delta\Lambda<0$ delays merger and extends the inspiral, increasing the available bandwidth and enhancing the gain. Together, these effects yield the observed monotonic, sub-quadratic scaling and the signed asymmetry in $\Delta\rho(\Delta\Lambda)$ seen in Fig.~\ref{fig:snr_scaling_signed} and Fig.~\ref{fig:snr_scaling_log}. We emphasize that the fitted power-law exponent ($\gamma \approx 1.24$) represents an empirical population average rather than a fundamental scaling constant. The substantial scatter ($R^2 \approx 0.55$) reflects that the SNR gain is not a function of $|\Delta\Lambda|$ alone, but is sensitive to the specific binary configuration, most notably the mass ratio and the precise number of cycles remaining after $t_{stitch}$, which varies stochastically across our injection set.
\end{enumerate}

\subsubsection{Softening vs. Stiffening Asymmetry}

While the magnitude of the phase shift depends on $|\Delta \Lambda|$, the sign of the transition introduces a physical asymmetry in the signal duration, leading to different recoverability profiles.

\begin{itemize}
    \item \textbf{Softening ($\Delta \Lambda < 0$):} A transition to a more compact EoS reduces the tidal drag, decelerating the frequency evolution ($\dot{f}$). This delays the merger, allowing the system to accumulate \textit{extra} phase cycles in the detector's most sensitive band compared to the null model. The \textsc{troye} model recovers this additional flux, leading to a higher potential SNR gain.
    \item \textbf{Stiffening ($\Delta \Lambda > 0$):} A transition to a less compact EoS increases tidal drag, accelerating the inspiral. Consequently, the merger occurs \textit{earlier} than predicted by the null model. The SNR gain here arises primarily from correctly truncating the template (avoiding the integration of noise after the merger), which yields a smaller net benefit than recovering missing signal cycles.
\end{itemize}

\subsection{Sensitivity and Distance Dependence}
\label{subsec:sensitivity_distance}

The ability to resolve a dynamic phase transition is naturally constrained by the signal-to-noise ratio (SNR), which scales inversely with the luminosity distance. Rather than a sharp cutoff, we observe a continuous degradation in detectability as the signal strength weakens. To quantify this gradient, we mapped the statistical support (measured via the natural log of the Bayes Factor, $\ln \mathcal{B}$) across the parameter space of distance $d_L$ and transition magnitude $|\Delta\Lambda|$.

To empirically verify the mechanics of this sensitivity loss, we performed a targeted stress test using a single reference binary with a large transition magnitude ($\Delta\Lambda \approx 1500$). We injected this identical signal at distances ranging strictly from $10$ to $1000$ Mpc. 

\begin{figure}[b]
    \centering
    \includegraphics[width=\columnwidth]{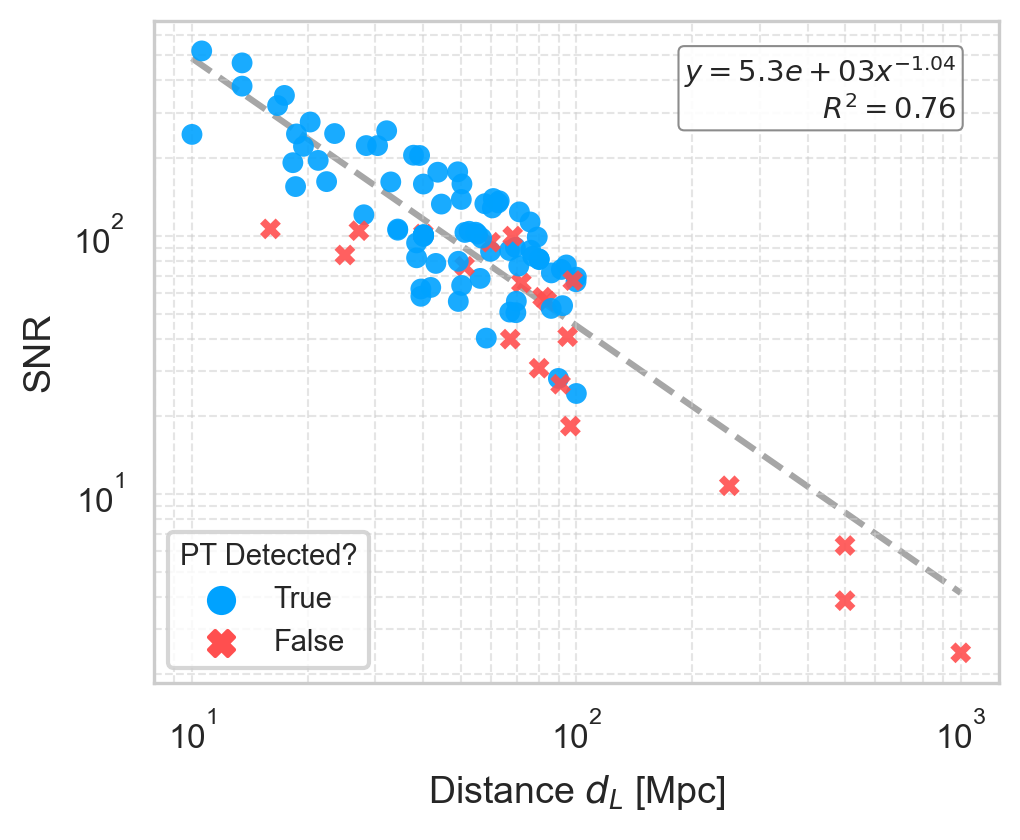}
    \caption{Degradation of parameter recovery with distance. The scatter points show the Matched Filter SNR for the simulated events. As the SNR drops, the ability to resolve the distinct $\Lambda_{pre}$ and $\Lambda_{post}$ values diminishes, eventually resulting in overlapping posteriors that mimic the prior distribution at large distances.}
    \label{fig:distance_scaling}
\end{figure}

As illustrated in Fig.~\ref{fig:distance_scaling}, we observe the expected inverse relationship between the optimal SNR and distance ($\rho \propto d_L^{-1}$). Consequently, the posterior distributions for the pre-transition ($\Lambda_{pre}$) and post-transition ($\Lambda_{post}$) tidal parameters—which are tightly constrained and disjoint at $10$ Mpc—progressively broaden. At $d_L > 250$ Mpc, the credible intervals begin to overlap significantly. By $1000$ Mpc, the parameter recovery effectively returns the prior distribution, rendering the transition indistinguishable from the null hypothesis despite the large injected magnitude.

\subsection{Stress Tests} \label{sec:stress_tests}
The baseline injection campaign utilized fixed extrinsic parameters to establish an upper bound on sensitivity. To validate the method under realistic conditions, we performed a series of ``stress tests'' with expanded prior volumes, introducing known degeneracies.

\begin{figure}[b]
    \centering
    \includegraphics[width=\columnwidth]{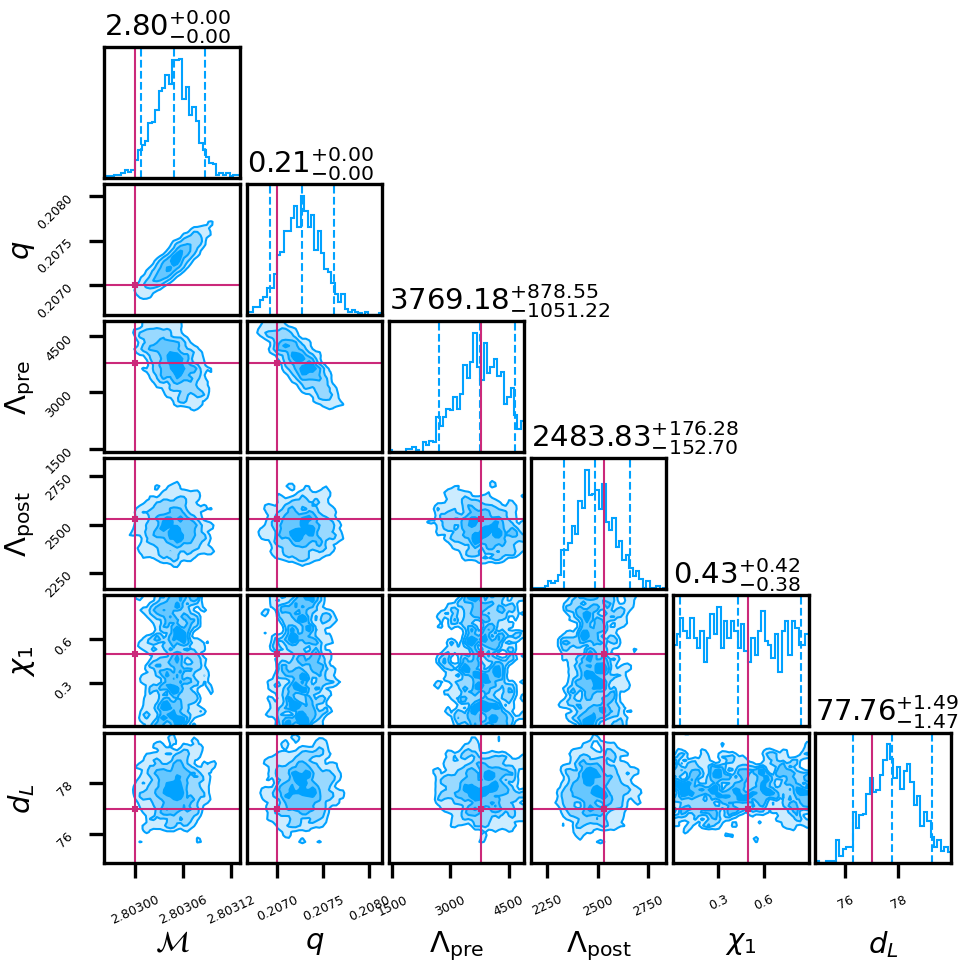}
    \caption{\label{fig:stress_test} Posterior distributions for a test event, where the mass ratio ($q$), luminosity distance ($d_L$), and black hole spin ($\chi_{BH}$) were free parameters. Despite the lack of constraint on the spin parameter (seen in the flat $\chi_1$ posterior), the pre- and post-transition tidal parameters ($\Lambda_{pre}, \Lambda_{post}$) remain well-separated, demonstrating that the phase transition signature is orthogonal to spin-orbit effects.}
\end{figure}

\subsubsection{Mass-Ratio--Tidal Correlations (Free $q$)}
For three representative injections, we relaxed the mass-ratio constraint and sampled uniformly over $q \in [0.133, 0.286]$. Although $q$ and the (constant) tidal deformability are known to be correlated in the inspiral phasing \cite{Flanagan2008}, this correlation does not imply an exact degeneracy in our transition search. The mass ratio influences the waveform phase and amplitude across the entire inspiral (starting at low PN order), and is therefore tightly constrained by the long-baseline signal content. In contrast, a phase transition produces a late-time, nonstationary imprint: $\Delta\Lambda$ modifies the phase evolution only after the transition (and can alter the effective duration relative to the stitching time), which cannot be replicated by a smooth change in $q$. Consistent with this, allowing $q$ to vary did not broaden the posterior on $\Delta\Lambda$, and the recovered Bayes factors remained comparable to the fixed-$q$ baseline, indicating that mass-ratio uncertainty does not mimic a transition signature within our prior range.

\subsubsection{Spin and Distance Correlations (Free $d_L,\chi_{BH}$)}
In a second subset of three events, we additionally sampled over the luminosity distance and the black-hole spin magnitude $\chi_{BH}$, increasing the intrinsic dimensionality of the inference problem. As expected for non-precessing systems at moderate SNR, $\chi_{BH}$ was only weakly constrained and largely reverted to the prior (Fig.~\ref{fig:stress_test}), whereas $d_L$ was tightly constrained and accurately recovered. Importantly, these parameters imprint the waveform in a qualitatively different way from a dynamical tidal transition: aligned-spin effects enter through smooth, secular phase corrections (spin--orbit and spin--spin terms) and $d_L$ primarily rescales the amplitude, whereas $\Delta\Lambda$ produces a localized late-inspiral change in the phase evolution tied to the transition time. Consistent with this morphological distinction, all three events still met our detection criteria, indicating that uncertainty in $(d_L,\chi_{BH})$ does not significantly degrade recovery of the transition signal in this regime.

\subsubsection{Temporal Resolution and Parameter Degeneracy}
\label{sec:t_stitch_recovery}
In the standard analysis, the time of the phase transition ($t_{stitch}$) was treated as a fixed parameter to maintain computational tractability. To assess the feasibility of measuring this parameter directly from the data, we conducted a series of stress tests where $t_{stitch}$ was relaxed and treated as a free parameter across three distinct configurations.

In a subset of two events, we first targeted the merger proximity, injecting the transition in the final moments of coalescence ($\sim 40$ ms before merger). The sampler successfully localized the transition epoch, identifying the approximate time of the event. However, due to the limited number of signal cycles remaining before merger, the recovered posterior distributions were relatively broad ($\sigma_t > 10$ ms).

\begin{figure}[h]
    \centering
    \includegraphics[width=\columnwidth]{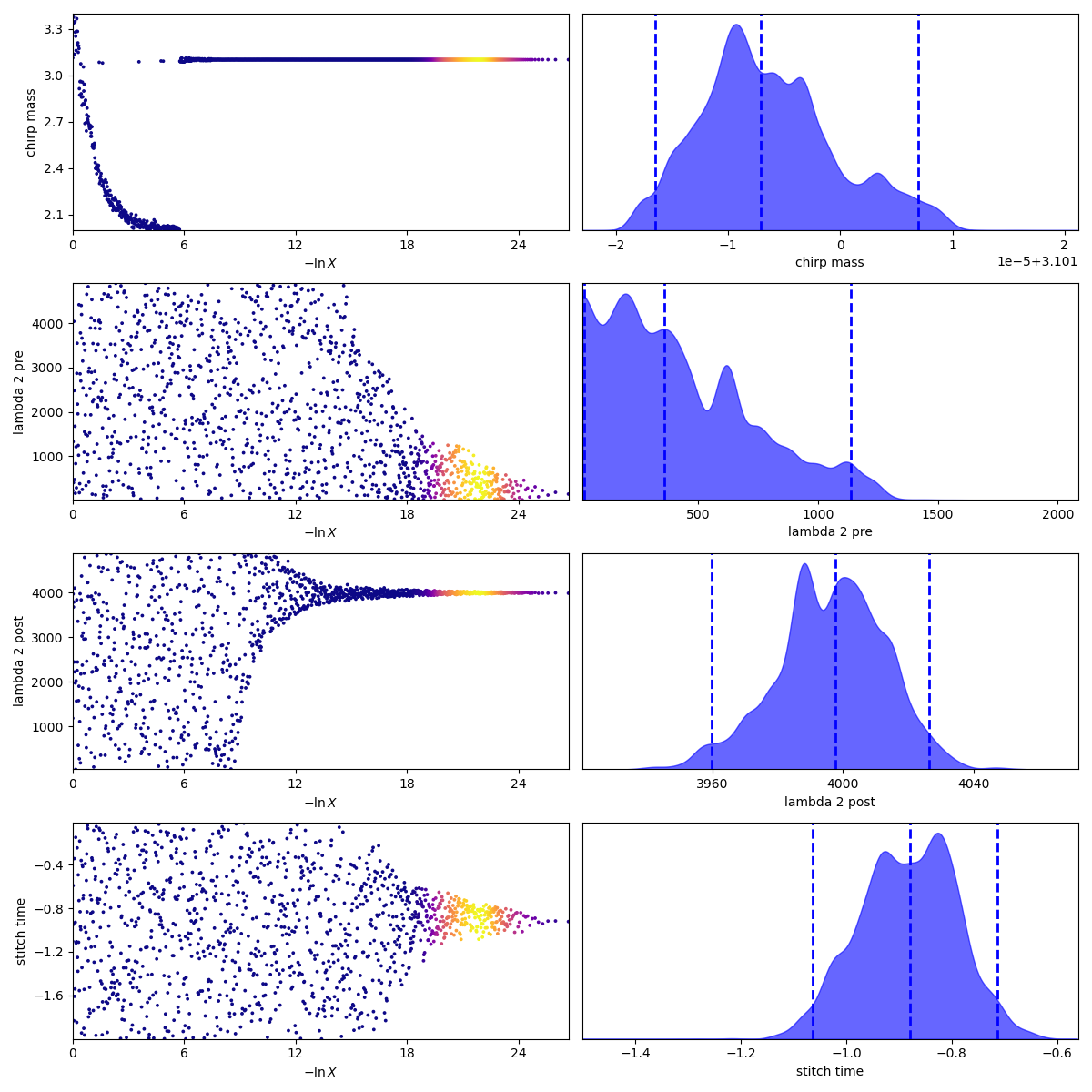}
    \caption{\label{fig:t_trace} Nested sampling trace plot for an event where the stitching time $t_{stitch}$ was a free parameter. The bottom-right panel shows the marginalized posterior for $t_{stitch}$, which successfully peaks around the injected value of $-0.8$s, rejecting the majority of the 2-second prior volume. This confirms that for sufficiently loud transitions, the timing of the event is recoverable.}
\end{figure}

Next, to test performance over larger temporal volumes, we analyzed transitions injected at earlier stages of the late inspiral ($-0.2$ s and $-0.8$ s relative to merger) using wide priors in a subset of two simulated events. As shown in Fig.~\ref{fig:t_trace}, the sampler successfully converged on the true values, excluding the majority of the prior volume. While the 90\% credible intervals remained broad ($>10$ ms), this result demonstrates that the algorithm can effectively search for transition candidates over second-long windows without getting trapped in local maxima.

Finally, we performed a full-degeneracy test on a subset of two events, simultaneously fitting for the Chirp mass $\mathcal{M}$, mass ratio $q$, tidal parameters ($\Lambda_{pre}, \Lambda_{post}$), and the transition time $t_{stitch}$. This represented the most complex test case. The recovery of the stitching time showed increased errors extending up to $\sim 20$ ms. However, crucially, the sampler successfully decoupled the timing uncertainty from the tidal magnitude. The posterior distributions for $\Lambda_{pre}$ and $\Lambda_{post}$ remained unaffected and well-separated, confirming that the \textit{presence} of the phase transition is robustly identified even when the exact timing is poorly constrained by degeneracies with the binary's mass parameters.

These results suggest that while pinpointing the exact millisecond of a phase transition is challenging due to the short duration of the tidal signal, the \textit{presence} and \textit{magnitude} of the transition are robust observables that persist even when the timing is treated as a free nuisance parameter.

\section{\label{sec:discussion}Discussion}

\subsection{Astrophysical Implications}
A dynamical phase transition during coalescence would provide a distinctive GW signature of exotic matter in NS cores. In our injection study, detectability is realistic for nearby systems ($d_L\!\lesssim\!100\,\mathrm{Mpc}$) if the transition produces a sizable tidal deformability shift, roughly $|\Delta\Lambda|\!\gtrsim\!400$ at design-level sensitivities. While this threshold implies a significant shift in the equation of state, it is consistent with hydrodynamic simulations of strong first-order phase transitions. For example, \cite{Prasad2018} reports shock-induced conversions that can drive a radial contraction of order $\Delta R\sim 1\,\mathrm{km}$. Since the tidal deformability scales steeply with compactness (approximately $\Lambda \propto R^5$ at fixed mass), a contraction from $R\simeq11.5$ km to $R\simeq10.5$ km would imply $\Lambda_{\rm after}/\Lambda_{\rm before}\simeq (10.5/11.5)^5 \approx 0.63$, i.e.\ a $\sim 40\%$ reduction—naturally large enough to yield $|\Delta\Lambda|$ of several hundred for typical pre-transition values.
The observability of such transitions is primarily SNR-limited. In our simulations, a single-detector SNR of $\rho\simeq 29$ yields $\sim 50\%$ detection efficiency for $|\Delta\Lambda|\!\gtrsim\!400$, while $\rho\simeq 65$ is required for $\sim 90\%$ efficiency (compared to $\rho\simeq13.9$ for GW200105, $\rho\simeq11.6$ for GW200115, and $\rho\simeq11.8$ for GW230529 \cite{abbott2021}\cite{abac2025gwtc}).
Nevertheless, even if individual events are only marginally informative, a hierarchical Bayesian analysis across an ensemble of NSBH detections could coherently accumulate evidence for (or against) a population-level preference for nonzero $|\Delta\Lambda|$, providing a viable path to testing this phenomenology with multiple sub-threshold candidates.

\subsection{Limitations and Future Work} \label{subsec:limitations}

This study assumes stationary Gaussian noise. While most tidal information accumulates over the long inspiral, the discriminating imprint of a dynamical transition is localized to the final $\sim 40\,\mathrm{ms}$ (roughly $200$--$450\,\mathrm{Hz}$ in our catalog). A glitch overlapping this time--frequency region can reduce sensitivity to $\Delta\Lambda$ and bias Bayes factors, especially if it affects multiple detectors or survives standard data-quality vetoes. A realistic assessment therefore requires injections into real O3/O4 strain together with standard mitigation (e.g., gating/vetoes and transient-noise marginalization), to quantify both detection efficiency and the false-positive rate in the presence of non-Gaussian noise.
Additionally, our model neglects neutron star spin and black hole precession. Precessional modulations can introduce phase evolution degeneracies that mimic or mask tidal effects. We also utilized a fixed transition window of $\tau = 10\,\mathrm{ms}$; treating $\tau$ as a free parameter in future inference would allow GWs to constrain the physical kinetics of the phase transition. Finally, while the \textsc{troye} framework is conceptually extensible to BNS systems, the phenomenology is significantly more complex. A comprehensive analysis would require modeling independent phase transitions for both stars, potentially occurring at distinct times ($t_{stitch,1} \neq t_{stitch,2}$). This increase in the dimensionality of the inference problem makes a robust quantitative survey of the BNS sector a subject for future study.

\section{Conclusion}
We have presented a framework for searching for dynamic phase transitions in NSBH coalescences. By focusing on NSBH systems, we avoid the ambiguities of BNS tidal extraction. The \textsc{troye} model provides a computationally efficient tool for this search. With the expected increase in NSBH detections in the upcoming O5 observing run, this method offers a new avenue to probe properties of matter at the most extreme densities in the universe.


\begin{acknowledgments}
This material is based upon work supported by NSF's LIGO Laboratory which is a major facility fully funded by the National Science Foundation.
This work has utilized computational resources provided by the LIGO Data Grid (LDG). We further acknowledge the use of high-performance workstations provided by Prof. Emanuele Dalla Torre and Prof. Jonathan Ruhman of Bar-Ilan University for the parameter estimation analysis.
O.~Dan has been supported by the US-Israel Binational Science Fund (BSF) grants No. 2020245 and 2024816, as well as the Israel Science Fund (ISF) grant No. 1698/22.
O.~Birnholtz has benefited from Israel's Council of Higher Education's Alon Scholarship.
\end{acknowledgments}

\clearpage

\bibliography{refs}

\end{document}